\documentclass[pdftex]{elsart}

\usepackage{graphicx, subfigure}
\usepackage{amssymb, amsfonts, amsmath}
\usepackage{hyperref, cite}
\usepackage[bf,footnotesize]{caption}

\newcommand{\Oa}{\mathcal{O}(a)}
\newcommand{\Oasq}{\mathcal{O}(a^2)}

\newcommand{\nft}{N_\mathrm{f}=2}

\newcommand{\nftoo}{N_\mathrm{f}=2+1+1}

\newcommand{\mpcac}{m_\mathrm{PCAC}}
\newcommand{\wsp}{W_6'}

\newcommand{\wep}{W_8'}
\newcommand{\mpc}{M_{\pi^\pm}}
\newcommand{\mpn}{M_{\pi^0}}
\newcommand{\mpnc}{M_{\pi^{(0,{\rm c})}}}
\newcommand{\re}{\operatorname{Re}}
\newcommand{\tr}{\operatorname{Tr}}
\newcommand{\Str}{\textrm{Str}}

\graphicspath{{figures/}}

\begin{document}

\begin{frontmatter}
  
  \vspace*{-1.0truecm} \title{Determination of Low-Energy Constants of
    Wilson Chiral Perturbation Theory}
  \vspace*{-0.5truecm}
  \begin{center}
    \includegraphics[draft=false,width=0.13\linewidth]{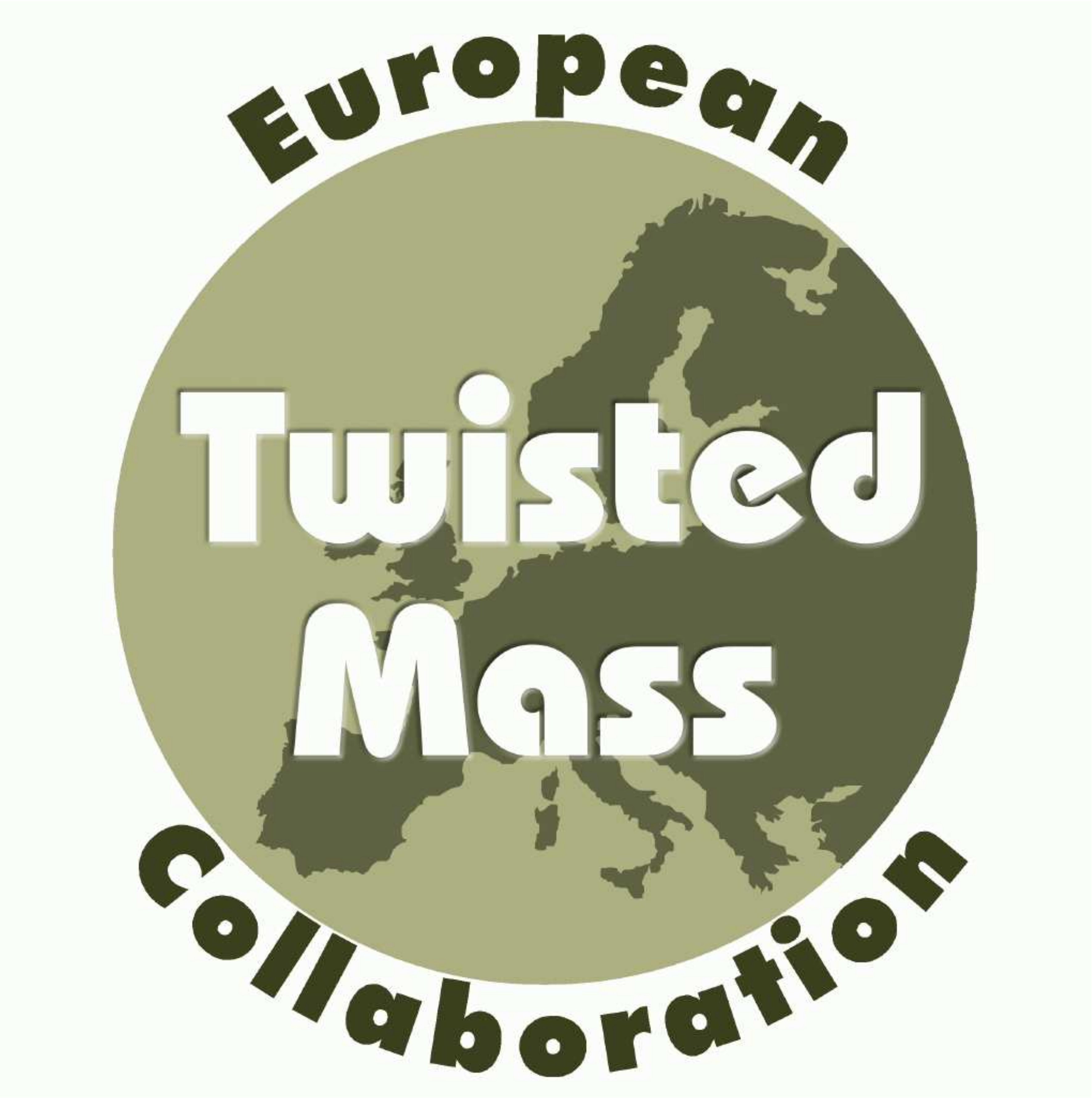}
  \end{center}

  \author[a,b]{Gregorio Herdo\'iza},
  \author[c,d]{Karl Jansen},
  \author[e]{Chris Michael},
  \author[f]{Konstantin Ottnad},
  \author[f]{Carsten Urbach}

  \address[a]{PRISMA Cluster of Excellence, Institut f{\"u}r Kernphysik\\
    Johannes Gutenberg-Universit{\"a}t, D-55099 Mainz, Germany}
  \address[b]{Departamento de F\'isica Te\'orica and Instituto de
    F\'isica Te\'orica UAM/CSIC,\\ Universidad Aut\'onoma de Madrid,
    Cantoblanco, E-28049 Madrid, Spain}
  \address[c]{NIC, DESY,
    Platanenallee 6, D-15738 Zeuthen, Germany}
  \address[d]{Department
    of Physics, University of Cyprus\\
    P.O.Box 20537, 1678 Nicosia,
    Cyprus}
  \address[e]{Theoretical Physics Division, Department of Mathematical
    Sciences\\ The University of Liverpool, Liverpool L69 3BX, UK}
  \address[f]{Helmholtz-Institut f{\"u}r Strahlen- und
    Kernphysik (Theorie) and Bethe Center for Theoretical Physics,
    Universit{\"a}t Bonn, 53115 Bonn, Germany}

  \begin{abstract}
    \noindent By matching Wilson twisted mass lattice QCD
    determinations of pseudoscalar meson masses to Wilson Chiral
    Perturbation Theory we determine the low-energy constants $\wsp$,
    $\wep$ and their linear combination $c_2$. We explore the
    dependence of these low-energy constants on the choice of the
    lattice action and on the number of dynamical flavours.
  \end{abstract}

  \begin{keyword}
    Lattice QCD, Wilson Fermions, Chiral Perturbation Theory.
    \PACS 12.38.Gc \sep 12.39Fe\\
    Preprint-No:~DESY 13-043,~FTUAM-13-127, IFT-UAM/CSIC-13-015,
    MITP/13-015, SFB/CPP-13-18\\
  \end{keyword}

\end{frontmatter}

%%%%%%%%%%%%%%%%%%%%%%%%%%%%%%%%%%%%%%%%%%%%%%%%%%%%%%%%%%%%%%%%%%%%%%%%%%%%%%
%%%%%%%%%%%%%%%%%%%%%%%%%%%%%%%%%%%%%%%%%%%%%%%%%%%%%%%%%%%%%%%%%%%%%%%%%%%%%%

\clearpage

\section{Introduction}

Lattice QCD simulations employing a discretisation of the Dirac
operator based on the original proposal by Wilson~\cite{Wilson:1974sk}
are currently being performed with light dynamical
fermions~\cite{DelDebbio:2006cn,Boucaud:2007uk,Lin:2008pr,Durr:2008zz,Aoki:2009ix,Baron:2010bv,Durr:2010aw,Bietenholz:2011qq}.
When decreasing the light quark mass at a fixed value of the lattice
spacing, a subtle interplay between mass and discretisation effects
can take place due to the explicit breaking of chiral symmetry by the
Wilson term.  In simulations with light values of the quark mass it
is, therefore, vital to understand and monitor the discretisation
effects and to obtain a quantitative measure of their size.

Close to the continuum limit, a useful way to determine the
discretisation effects in the regime of light quark masses is provided
by Wilson chiral perturbation theory (W$\chi$PT), an extension of the
continuum chiral effective theory including additional terms
proportional to powers of the lattice
spacing~\cite{Sharpe:1998xm}. Depending on the order of the expansion,
additional low energy constants (LECs) appear, whose values are not
known a priori: they depend on the lattice action and can only be
computed from a simulation.

Knowing the values of the LECs of W$\chi$PT is of particular interest,
because W$\chi$PT predicts a non-trivial phase structure for Wilson
type fermions in the lattice spacing and quark mass
plane~\cite{Aoki:1984qi,Creutz:1996bg,Sharpe:1998xm}. Depending on the
sign of a particular combination of Wilson LECs -- commonly denoted as
$c_2\propto -(2\wsp +\wep)$ -- either the
Aoki-scenario~\cite{Aoki:1984qi} or a first order, so called
Sharpe-Singleton~\cite{Sharpe:1998xm} scenario is realised. Numerical
evidence for both scenarios has been observed in lattice QCD
simulations and dedicated studies of the associated phase diagrams
have been performed by several
groups~\cite{Aoki:1992nb,Blum:1994eh,Aoki:1995ft,Aoki:1996af,Aoki:2001xq,Ilgenfritz:2003gw,Aoki:2004iq,Farchioni:2004us,Farchioni:2004fs,Farchioni:2005tu,Farchioni:2005bh,Chiarappa:2006ae}.

The Aoki scenario with a positive $c_2$ was found to be realised in
quenched simulations.  In dynamical simulations the Sharpe-Singleton
scenario with negative $c_2$ was observed when using Wilson twisted
mass fermions at maximal
twist~\cite{Farchioni:2004us,Farchioni:2004fs,Farchioni:2005tu,Farchioni:2005bh}. This
manifests itself in the fact that the neutral pion mass $\mpn$ is
lighter than the charged one, $\mpc$, where the splitting of the
squared masses is proportional to $c_2$. In turn, a measurement of the
pion mass-splitting in Wilson twisted mass lattice QCD provides a way
to measure $c_2$ and, hence, the LECs of the corresponding chiral
effective theory.  However, this way of computing $c_2$ is challenging
since for the neutral pion mass disconnected contributions need to be
evaluated.

There are also alternative ways to determine the LECs of W$\chi$PT. In
Refs.~\cite{Deuzeman:2011dh,Damgaard:2011eg,Damgaard:2012gy,Damgaard:2013xi}
they have been studied by matching the analytical
predictions~\cite{Damgaard:2010cz,Akemann:2010em,Splittorff:2011bj,Splittorff:2011zk,Splittorff:2012gp,Kieburg:2012fw,Damgaard:2012gy}
for the spectrum of the Wilson Dirac operator -- with fixed index in a
finite volume -- to lattice data.\footnote{For a recent review, we
  refer to~\cite{Splittorff:2012hz}.} Determinations of the Wilson
LECs have also been carried out via the spectral density of the
Hermitian Wilson-Dirac
operator~\cite{Necco:2011vx,Necco:2011jz,Necco:2013sxa}. Lattice
determinations of the pion scattering lengths have been used in
Refs.~\cite{Aoki:2008gy,Bernardoni:2011fx}, an approach that was
extended to a partially quenched setup in Ref.~\cite{Hansen:2011mc}.
In a mixed action with Wilson-type sea fermions and chirally invariant
valence quarks, a mixed action chiral
Lagrangian~\cite{Bar:2002nr,Bar:2003mh} can be constructed. The
corresponding LECs -- in particular $\wep$ -- have been recently
determined in the case of overlap valence quarks on a Wilson twisted
mass sea~\cite{Cichy:2012vg}.

In this paper we are going to determine the LECs using a method
recently proposed in Ref.~\cite{Hansen:2011kk}. It relies on the
measurement of pseudoscalar meson masses involving Wilson twisted mass
fermions. In this approach, the LEC $\wep$ is related to the
mass-splitting between the charged pion mass $\mpc$ and the
``connected neutral pion'' mass $\mpnc$. The latter is determined from the
quark-connected correlation, which contributes to the neutral pion in
twisted mass QCD. The ``connected neutral pion'' correlation function thus differs from the
complete correlation function needed to determine the neutral pion by
the absence of disconnected diagrams. This has the advantage that in
numerical studies the ``connected pion mass-splitting'' should be
accessible with good statistical precision. In addition, the
mass-splitting, $\mpnc^2 - \mpn^2$, between the connected and full
neutral pion mass, provides an estimate of $\wsp$. We are also going
to study the mass and lattice spacing dependence of these splittings
arising at higher order in the W$\chi$PT expansion.

The paper is structured as follows: in the next section we collect the
W$\chi$PT expressions relating the pion mass-splittings to the Wilson
LECs. In Section~\ref{sec:actions} we present the lattice actions used
in our study. The determination of the LECs from two of those lattice
setups is described in Section~\ref{sec:num}. Finally, a qualitative
comparison of the values of these LECs from different choices of the
lattice action is reported in Section~\ref{sec:discussion}.

%%%%%%%%%%%%%%%%%%%%%%%%%%%%%%%%%%%%%%%%%%%%%%%%%%%%%%%%%%%%%%%%%%%%%%%%%%%%%%

\section{Wilson Chiral Perturbation Theory (W$\chi$PT)}

In this section, we briefly discuss Wilson chiral perturbation
theory and introduce the expressions used in our study. For a recent
review on the applications of $\chi$PT to lattice QCD, we refer to
Ref.~\cite{Golterman:2009kw}.

Our study will be based on the computation of pseudoscalar meson
masses involving Wilson twisted mass fermions. We are therefore
interested in a chiral Lagrangian involving a mass matrix $M =
m_0^{\rm R} + i\mu_\ell^{\rm R}\tau_3$, where $m_0^{\rm R}$ and
$\mu_\ell^{\rm R}$ are the renormalised untwisted and twisted quark
masses, respectively. The masses $m_0$ and $\mu_\ell$ appear in the
Wilson twisted mass action of eq.~(\ref{eq:sl}).

At leading order (LO) in the power counting, $m_0\sim \mu_\ell\sim
a^2\Lambda_{\rm QCD}^3$, and after a shift of the quark mass to remove
a term of ${\cal O}(a)$, the partially quenched chiral Lagrangian
reads~\cite{Sharpe:1998xm,Bar:2003mh}
%%%%%%%%%%%%%%%%%%%%%%%%%%%%
\begin{eqnarray}
  {\cal L}_\chi &=&
  \frac{f^2}{8} \,\Str\left(
  \partial_\mu \Sigma \partial_\mu \Sigma^\dagger\right)
  -
  \frac{f^2\,B_0}{4} \,\Str\left( M^\dagger \Sigma+\Sigma^\dagger M\right)
  \nonumber\\
  && - \hat{a}^2 W_6' \,\left[\Str\left(\Sigma+\Sigma^\dagger\right)\right]^2
  - \hat{a}^2 W_7' \,\left[\Str\left(\Sigma-\Sigma^\dagger\right)\right]^2
  \nonumber\\
  &&
  - \hat{a}^2 W_8' \,\Str\left(\Sigma^2+[\Sigma^\dagger]^2\right)
  \,,
  \label{eq:LPQWChPT}
\end{eqnarray}
%%%%%%%%%%%%%%%%%%%%%%%%%%%%
where $\Sigma$ parametrises the vacuum manifold and thus characterises
the Nambu-Goldstone bosons arising from the spontaneous breaking of
chiral symmetry. The traces over the flavour indices are denoted by
Str and $\hat a = 2 W_0 a$. In addition to the continuum LECs $B_0$
and $f$ (defined with the convention giving $f_\pi\approx 130\;$MeV),
the chiral Lagrangian also includes $W_0$, $\wsp$, $W'_7$ and $\wep$,
which are Wilson LECs describing discretisation effects.

In this work we are interested in the determination of the Wilson LECs
by matching lattice QCD calculations of pseudoscalar meson masses to
their PQW$\chi$PT expressions. As already mentioned, we consider
Wilson twisted mass fermions at maximal twist. This is achieved in the
chiral Lagrangian by setting $m_0=0$. At non vanishing values of the
lattice spacing, the breaking of flavour symmetry by the twisted mass
term in eq.~(\ref{eq:sl}) implies that the mass of the charged pion
$\mpc$ differs from that of the neutral pion $\mpn$ by O($a^2$)
effects. A similar pattern holds for the mass of the ``connected
neutral pion'' mass, $\mpnc$. The PQW$\chi$PT expressions for these three
meson masses at LO
read~\cite{Munster:2004am,Scorzato:2004da,Sharpe:2004ps,Hansen:2011kk},
%%%%%%%%%%%%%%%%%%%%%%%%%%%%
\begin{eqnarray}
  \mpc^2 \ &=&\ 2 B_0 \mu_\ell \,,
  \label{eq:mpc}
  \\
  \mpn^2 \ &=&\ 2 B_0 \mu_\ell - 8a^2 \,(2 w'_6 + w'_8) \,,
  \label{eq:mpn}
  \\
  \mpnc^2 \ &=&\  2 B_0 \mu_\ell - 8a^2\,w'_8
  \,,
  \label{eq:mpinc}
\end{eqnarray}
%%%%%%%%%%%%%%%%%%%%%%%%%%%%
where $w'_{k}$ is related to the Wilson LEC  $W'_{k}$ by
%%%%%%%%%%%%%%%%%%%%%%%%%%%%
\begin{eqnarray}
  w'_k \ =\  \frac{16 W_0^2\,W'_k}{f^2} \qquad (k=6,8)
  \,. \label{eq:wk}
\end{eqnarray}
%%%%%%%%%%%%%%%%%%%%%%%%%%%%

To determine the individual values of the LECs it is, therefore,
possible to consider the following mass-splittings:
%%%%%%%%%%%%%%%%%%%%%%%%%%%%
\begin{eqnarray}
  \mpc^2 - \mpnc^2 \  &=& \  8a^2\, w'_8\,,  \label{eq:w8ms}\\
  \frac{1}{2} \, \left( \mpnc^2 - \mpn^2 \right) \  &=& \  8a^2\, w'_6\,. \label{eq:w6ms}
\end{eqnarray}
%%%%%%%%%%%%%%%%%%%%%%%%%%%%

From eq.~(\ref{eq:mpn}), it appears that the linear combination of
LECs which controls the mass-splitting between charged and neutral
pions is given by
%%%%%%%%%%%%%%%%%%%%%%%%%%%%
\begin{equation}
  c_2 \ = \ -\frac{32 W_0^2}{f^2}(2 \wsp + \wep)\,.
  \label{eq:c2}
\end{equation}
%%%%%%%%%%%%%%%%%%%%%%%%%%%%
This can also be re-expressed as
%%%%%%%%%%%%%%%%%%%%%%%%%%%%
\begin{eqnarray}
  c_2 \ &=& \ \frac{1}{4a^2}\, \left( \mpn^2 - \mpc^2 \right) \,, 
  \label{eq:c2ms}\\ ~ \nonumber \\
  &=& \ -2 \, ( 2 w'_6 + w'_8) \,.
  \label{eq:c2w}
\end{eqnarray}
%%%%%%%%%%%%%%%%%%%%%%%%%%%%

In the W$\chi$PT expressions presented above, two light
mass-degenerate flavours were assumed to be present in the sea
sector. When considering also other dynamical flavours, such as the
strange and the charm quarks, the same expressions hold when assuming
that these heavier flavours sufficiently decouple from the light quark
sector. In this case, the values of the Wilson LECs will have a
further residual dependence on the heavier quark masses.

Before closing this section, we mention that W$\chi$PT calculations at
NLO have been carried out for the pion mass and decay constant with
$\nft$~\cite{Colangelo:2010cu,Bar:2010jk,Ueda:2011ib} and
$\nftoo$~\cite{Munster:2011gh} flavours of twisted mass fermions.

%%%%%%%%%%%%%%%%%%%%%%%%%%%%%%%%%%%%%%%%%%%%%%%%%%%%%%%%%%%%%%%%%%%%%%%%%%%%%%
%%%%%%%%%%%%%%%%%%%%%%%%%%%%%%%%%%%%%%%%%%%%%%%%%%%%%%%%%%%%%%%%%%%%%%%%%%%%%%

\section{Lattice actions}
\label{sec:actions}

The complete lattice action can be written as
%%%%%%%%%%%%%%%%%%%%%%%%%%%%
\begin{equation}
  S= S_f + S_g\, ,
\end{equation}
%%%%%%%%%%%%%%%%%%%%%%%%%%%%
where $S_f$ is the fermionic action and $S_g$ is the pure gauge
action. As we shall see, in this work we will consider a few
alternatives for both the fermionic and the gauge actions in order to
explore the dependence of the Wilson LECs on the details of the
lattice action.  As discussed below, we will use for $S_F$ a few
variants of Wilson twisted mass fermions.

%%%%%%%%%%%%%%%%%%%%%%%%%%%%%%%%%%%%%%%%%%%%%%%%%%%%%%
%%%%%%%%%%%%%%%%%%%%%%%%%%%%%%%%%%%%%%%%%%%%%%%%%%%%%%

\subsection{Wilson Twisted Mass Fermions}

The Wilson twisted mass (Wtm) lattice action for the mass degenerate
light doublet $(u,d)$ in the so-called twisted basis
reads~\cite{Frezzotti:2000nk,Frezzotti:2003ni},
%%%%%%%%%%%%%%%%%%%%%%%%%%%%
\begin{equation}
  \label{eq:sl}
  S_l\ =\ a^4\sum_x\left\{ \bar\chi_l(x)\left[ D[U] + m_{0,l} +
    i\mu_\ell\gamma_5\tau_3\right]\chi_l(x)\right\}\, ,
\end{equation}
%%%%%%%%%%%%%%%%%%%%%%%%%%%%
where $m_{0,l}$ is the untwisted bare quark mass, $\mu_\ell$ is the bare
twisted light quark mass, $\tau_3$ is the third Pauli matrix acting in
flavour space and
%%%%%%%%%%%%%%%%%%%%%%%%%%%%
\begin{equation}
  \label{eq:DW}
  D[U] = \frac{1}{2}\left[\gamma_\mu\left(\nabla_\mu +
    \nabla^*_\mu\right) -a\nabla^*_\mu\nabla_\mu \right]\,,
\end{equation}
%%%%%%%%%%%%%%%%%%%%%%%%%%%%
is the massless Wilson-Dirac operator. $\nabla_\mu$ and $\nabla^*_\mu$
are the forward and backward gauge covariant difference operators,
respectively. Twisted mass light fermions are said to be at maximal
twist if the bare untwisted mass $m_{0,l}$ is tuned to its critical
value, $m_{\rm crit}$. The quark doublet $\chi_l=(\chi_u, \chi_d)$ in
the twisted basis is related by a chiral rotation to the quark doublet
in the physical basis
%%%%%%%%%%%%%%%%%%%%%%%%%%%%
\begin{equation}
  \psi_{l}^{phys} = e^{\frac{i}{2}\omega_{l}\gamma_{5}\tau_{3}} \, \chi_{l}, 
  \qquad \bar{\psi}_{l}^{phys}=\bar{\chi}_{l} \, e^{\frac{i}{2}\omega_{l}\gamma_{5}\tau_{3}}\, ,
\end{equation}
%%%%%%%%%%%%%%%%%%%%%%%%%%%%
where $\omega_{l}$ is the twist angle.

The twisted mass parameter $\mu_\ell$ provides an infrared regulator
avoiding the presence of accidental zero-modes in the Wilson-Dirac
operator. An important property of Wtm fermions is that at maximal
twist physical observables are O($a$)
improved~\cite{Frezzotti:2003ni}. In numerical simulations, maximal
twist is achieved by tuning the value of the hopping parameter
$\kappa=1/(2m_{0,l}+8)$ to its critical value $\kappa_{\rm crit}$ by
tuning the PCAC quark mass $\mpcac$ to zero. The expected O($a^2$)
scaling of physical observables when performing the continuum limit
extrapolation has been confirmed in the quenched
approximation~\cite{Jansen:2003ir,Jansen:2005gf,Jansen:2005kk,Abdel-Rehim:2005gz}
and with
$\nft$~\cite{Urbach:2007rt,Dimopoulos:2007qy,Alexandrou:2008tn,Baron:2009wt}
and $\nftoo$~\cite{Baron:2010bv,Drach:2010hy,Herdoiza:2011gp}
dynamical quarks.

A peculiar lattice artifact can appear in observables made out of Wtm
quarks due to the breaking of isospin and parity by the twisted mass
term in eq.~(\ref{eq:sl}). This effect, which is expected to vanish in
physical quantities at a rate of O($a^2$) when approaching the
continuum limit, has been observed to be numerically small in most of
the observables which have been
analysed~\cite{Urbach:2007rt,Dimopoulos:2007qy,Alexandrou:2008tn,Baron:2009wt,Baron:2010bv,Drach:2010hy}

An exception to this observed small isospin breaking effects is found
in the case of the neutral pseudoscalar mass. Indeed, isospin breaking
induces a mass-splitting between charged and neutral pion
masses. Dedicated numerical studies indicate that while in the charged
pion mass only very mild cutoff effects are present, the neutral pion
mass is instead affected by significant $\Oasq$
effects~\cite{Jansen:2005cg,Urbach:2007rt,Baron:2009wt,Baron:2010bv,Herdoiza:2011gp}. An
analysis based on the Symanzik expansion indicates that isospin
breaking affects only a limited set of observables in a sizeable way,
namely the neutral pion mass and kinematically related
quantities~\cite{Frezzotti:2007qv,Dimopoulos:2009qv}. This analysis is
complementary to that based on W$\chi$PT where, as previously
mentioned, the mass-splitting between charged and neutral pions is
parametrised by the combination of LECs appearing in $c_2$ defined in
eq.~(\ref{eq:c2}).

The determination of the neutral pion mass $\mpn$ involves both
connected and disconnected contributions. The computation of
quark-disconnected diagrams is challenging and requires the employment
of specific techniques in order to achieve a statistically significant
determination of $\mpn$~\cite{Michael:2007vn,Boucaud:2008xu}.

\subsubsection*{$\nftoo$ Wtm fermions}

In addition to a doublet of mass-degenerate light quarks ($u$,$d$), a
heavier doublet with strange and charm quarks -- ($s$,$c$) -- can be
incorporated in lattice QCD studied with Wilson twisted mass
fermions~\cite{Frezzotti:2004wz,Frezzotti:2003xj}. Also in this
$\nftoo$ setup, a tuning to maximal twist by imposing $\mpcac=0$,
allows to achieve the automatic O($a$) improvement of physical
observables. We refer to Ref.~\cite{Baron:2010bv} for a complete
description of the lattice action for the heavier quark doublet and
for further details on the lattice setup.

\subsubsection*{Mixed action with Osterwalder-Seiler valence quarks}

Osterwalder-Seiler (OS) valence quarks~\cite{Osterwalder:1977pc} can
be viewed as the building blocks of Wilson twisted mass fermions at
maximal twist. The OS action for an {\it individual} quark flavour
$\chi_f$ reads
%%%%%%%%%%%%%%%%%%%%%%%%%%%%
\begin{equation}
  \label{eq:os}
  S_f^\mathrm{OS}\ =\ a^4\sum_x\left\{ \bar\chi_f(x)\left[ D[U] + m_\mathrm{crit} +
    i\mu_f\gamma_5 r_f\right]\chi_f(x)\right\}\, ,
\end{equation}
%%%%%%%%%%%%%%%%%%%%%%%%%%%%
where $r_f$ (here $|r_f|=1$) is the Wilson parameter and $m_{\rm
  crit}$ the critical mass. By combining two flavours of OS quarks
with opposite signs of $r_f$, e.g. $r_2 = -r_1$, the action of a
doublet of maximally Wtm fermions of mass $\mu_f=\mu_1=\mu_2$ can be
recovered. The benefits of the OS action are that $\Oa$ improved
physical observables~\cite{Frezzotti:2004wz} can be obtained by using
the same estimates of $m_{\rm crit}$ as in the Wtm case and, thus,
avoiding further tuning effort. OS and Wtm fermions coincide with
Wilson fermions in the massless limit and consequently share the same
renormalisation factors. This simplifies the matching of sea and
valence quark masses in the context of a mixed-action with Wtm sea and
OS valence quarks.

The pseudoscalar correlation function obtained when considering only
the connected 
contribution to the neutral pion correlation function ({\it i.e.} when
ignoring disconnected diagrams), is precisely the pion correlator
with OS fermions. The mass-splittings in
eqs.(\ref{eq:w8ms})-(\ref{eq:w6ms}) can hence be interpreted as
involving sea and valence quarks in a mixed action setup.

In this work, we aim at determining the Wilson LECs in a lattice
theory with $\nftoo$ Wtm fermions. A particular effort will be devoted
to addressing the main systematic effects present in these
determinations. We furthermore aim at exploring the qualitative change
on the values of these LECs when varying the details of the lattice
action. Below, we briefly summarise the alternative lattice setups
used in this work. We will consider variants of the action differing
by the presence of smearing of the gauge links in the covariant
derivative, the inclusion of the Sheikholeslami-Wohlert term or by a
change in the number of dynamical flavours.

\subsubsection*{Stout Smearing.}

A smearing procedure can be applied to the gauge links entering in the
covariant derivatives in eq.~(\ref{eq:DW}). The stout
smearing~\cite{Morningstar:2003gk} procedure is analytic in the
un-smeared link variables and hence well suited for simulations with
the HMC algorithm. The smearing can be iterated several times, with
the price of extending the coupling of fermions to the gauge links
over a larger region.

\subsubsection*{Sheikholeslami-Wohlert term.}

In our comparison of the values of the mass-splittings in
eqs.~(\ref{eq:w8ms})-(\ref{eq:w6ms}) from different lattice setups, we
will also consider results available in the literature from quenched
lattice simulations with Wtm quarks including the
Sheikholeslami-Wohlert term~\cite{Sheikholeslami:1985ij}.

%%%%%%%%%%%%%%%%%%%%%%%%%%%%%%%%%%%%%%%%%%%%%%%%%%%%%%
%%%%%%%%%%%%%%%%%%%%%%%%%%%%%%%%%%%%%%%%%%%%%%%%%%%%%%

\subsection{Gauge action}
\label{sec:gauge}

The lattice gauge actions considered in this work have a generic form
which includes a plaquette term $U^{1\times1}_{x,\mu,\nu}$ and
rectangular $(1\times2)$ Wilson loops $U^{1\times2}_{x,\mu,\nu}$,
%%%%%%%%%%%%%%%%%%%%%%%%%%%%
\begin{equation}
  \label{eq:Sg}
  S_g =  \frac{\beta}{3}\sum_x\Biggl(  b_0\sum_{\substack{
      \mu,\nu=1\\1\leq\mu<\nu}}^4\{1-\re\tr(U^{1\times1}_{x,\mu,\nu})\}\Bigr. 
  \Bigl.\,+\,
  b_1\sum_{\substack{\mu,\nu=1\\\mu\neq\nu}}^4\{1
  -\re\tr(U^{1\times2}_{x,\mu,\nu})\}\Biggr)\, ,
\end{equation}
%%%%%%%%%%%%%%%%%%%%%%%%%%%%
with $\beta=6/g_0^2$ the bare inverse coupling and the normalisation
condition $b_0=1-8b_1$. We will consider the case of the Wilson
plaquette~\cite{Wilson:1974sk} action ($b_1=0$), the tree-level
Symanzik improved~\cite{Weisz:1982zw,Weisz:1983bn} action
($b_1=-1/12$) and the
Iwasaki~\cite{Iwasaki:1985we,Iwasaki:1996sn,Iwasaki:2011jk}
action~($b_1=-0.331$).  In Wtm simulations, the strength of the phase
transition has been found~\cite{Farchioni:2004fs,Farchioni:2005tu} to
depend on the value of the parameter $b_1$ in eq.~(\ref{eq:Sg}).

%%%%%%%%%%%%%%%%%%%%%%%%%%%%%%%%%%%%%%%%%%%%%%%%%%%%%%%%%%%%%%%%%%%%%%%%%%%%%%
%%%%%%%%%%%%%%%%%%%%%%%%%%%%%%%%%%%%%%%%%%%%%%%%%%%%%%%%%%%%%%%%%%%%%%%%%%%%%%

\section{Numerical Studies}
\label{sec:num}

%%%%%%%%%%%%%%%%%%%%%%%%%%%%%%%%%%%%%%%%%%%%%%%%%%%%%%
%%%%%%%%%%%%%%%%%%%%%%%%%%%%%%%%%%%%%%%%%%%%%%%%%%%%%%

\subsection{$\nftoo$ Wtm fermions with Iwasaki gauge action}

The purpose of this study is to determine the mass-splittings in
eqs.~(\ref{eq:w8ms}) and ~(\ref{eq:w6ms}), which are directly related
to the Wilson LECs $\wep$ and $\wsp$, respectively. The lattice action
is composed of the Iwasaki gauge action and $\nftoo$ flavours of
Wilson twisted mass fermions.

The simulations~\cite{Baron:2010bv} were performed at three values of
the lattice gauge coupling, $\beta =1.90$, $1.95$ and $\beta=2.10$,
corresponding to values of the lattice spacing $a\approx 0.09$\,fm,
$0.08$\,fm and $0.06$\,fm, respectively. The charged pion mass $\mpc$
approximately ranges from $230$\,MeV to $510$\,MeV.  Simulated volumes
correspond to values of $\mpc L$ larger than $3.3$. Physical spatial
volumes range from $(1.9\,\mathrm{fm})^3$ to $(2.8\,\mathrm{fm})^3$.

\begin{figure}[t!]
  \centering
  \subfigure[\label{fig:MS_con_211a}]{
    \includegraphics[height=6.5cm]{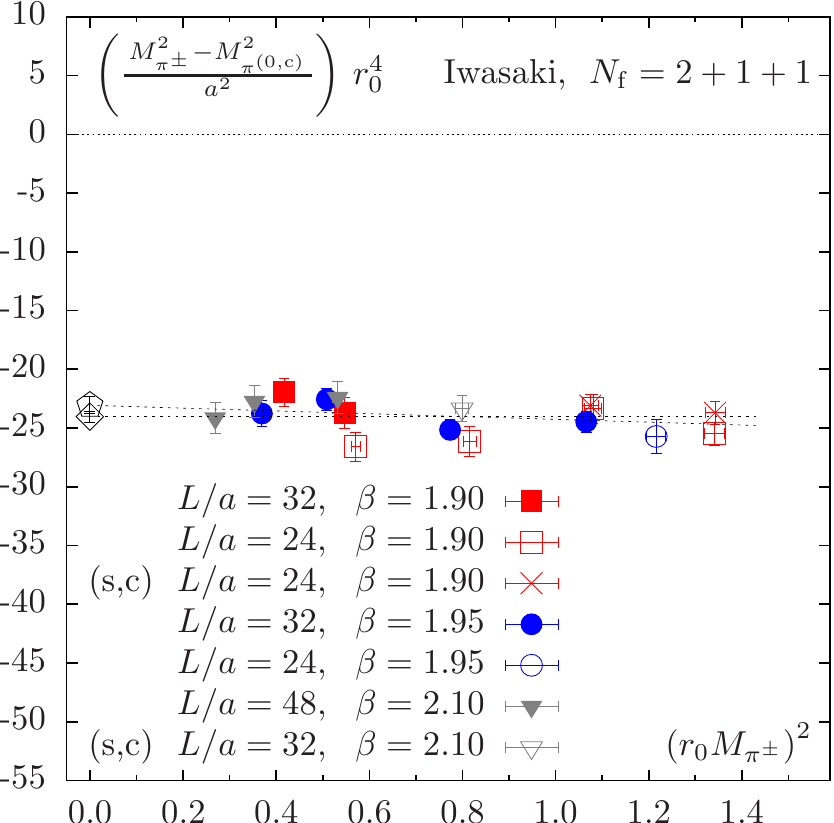}
  }
  \subfigure[\label{fig:MS_con_211b}]{
    \includegraphics[height=6.5cm]{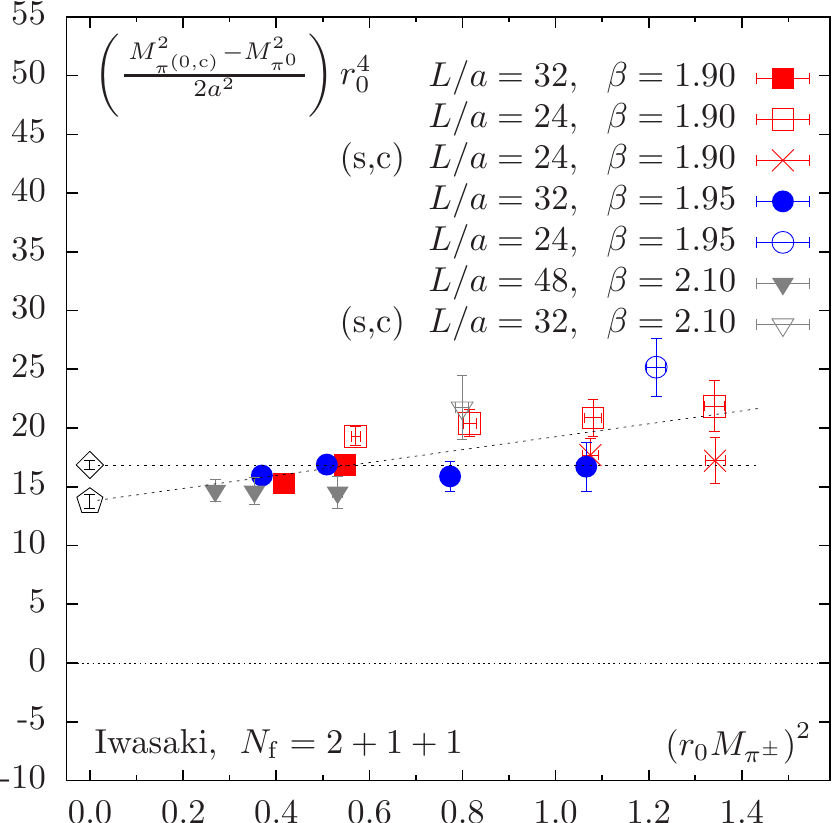}
  }
  \caption{Determination of the mass-splittings (a) $(\mpc^2 - \mpnc^2)/a^2$
    in eq.~(\ref{eq:w8ms}) and (b) $(\mpnc^2 - \mpn^2)/2a^2$ in
    eq.~(\ref{eq:w6ms}) as a function of $\mpc^2$. These quantities
    are in units of the chirally extrapolated Sommer scale $r_0$. The
    mass-splittings are directly related to the Wilson LECs $\wep$ and
    $\wsp$. The lattice setup with the Iwasaki gauge action and
    $\nftoo$ flavours of Wilson twisted mass fermions is
    considered. Filled and empty symbols signal a change in the
    lattice size. The label ``(s,c)'' in the legend indicates the
    effect of changing the values of the strange and charm quark
    masses in the sea. The mass-splittings illustrated in these
    figures assess the size of O($a^2$) discretisation effects. Hence,
    any lattice spacing dependence in these quantities points to
    residual higher order discretisation effects. The results of a
    chiral extrapolation from a constant and a linear fit in $\mpc^2$
    are shown. The deviation between the two extrapolated values is
    included in the systematic error analysis.}
  \label{fig:MS_con_211}
\end{figure}

The values of the pseudoscalar meson masses for the $\nftoo$ ensembles
are collected in Table~\ref{tab:nf211}.
%%%%%%%%%%%%%%%%%%%%%%%%%%%%%%%%%%
\begin{table}[t!]
  \centering
  \begin{tabular*}{1.0\textwidth}{@{\extracolsep{\fill}}lccccccc}
    \hline\hline
    Ens. & $\beta$ & $L/a$ & $a\mu_\ell$ & $a\mpc$ & $a\mpnc$ & $a\mpn$ & $r_0/a$\\
    \hline
    A30.32   & 1.90 & 32 & 0.0030 & 0.1234(03) & 0.2111(33) & 0.0611(036) & 5.23(4)\\
    A40.32   &      &    & 0.0040 & 0.1415(04) & 0.2274(31) & 0.0811(050) & \\
    A40.24   &      & 24 & 0.0040 & 0.1445(06) & 0.2375(25) & 0.0694(065) & \\
    A60.24   &      &    & 0.0060 & 0.1727(06) & 0.2544(26) & 0.1009(113) & \\
    A80.24   &      &    & 0.0080 & 0.1987(06) & 0.2659(25) & 0.1222(157) & \\
    A100.24  &      &    & 0.0100 & 0.2215(04) & 0.2883(14) & 0.1570(178) & \\
    \hline
    A80.24s  & 1.90 & 24 & 0.0080 & 0.1982(04) & 0.2649(16) & 0.1512(115) & \\
    A100.24s &      &    & 0.0100 & 0.2215(04) & 0.2841(16) & 0.1863(141) & \\
    \hline
    B25.32   & 1.95 & 32 & 0.0025 & 0.1064(07) & 0.1836(21) & 0.0605(036) & 5.71(4)\\
    B35.32   &      &    & 0.0035 & 0.1249(07) & 0.1919(17) & 0.0710(061) & \\
    B55.32   &      &    & 0.0055 & 0.1540(04) & 0.2177(19) & 0.1323(080) & \\
    B75.32   &      &    & 0.0075 & 0.1808(05) & 0.2360(12) & 0.1557(126) & \\
    B85.24   &      & 24 & 0.0085 & 0.1931(08) & 0.2480(11) & 0.1879(180) & \\
    \hline
    D15.48   & 2.10 & 48 & 0.0015 & 0.0695(03) & 0.1124(15) & 0.0561(031) & 7.46(6)\\
    D20.48   &      &    & 0.0020 & 0.0797(05) & 0.1170(16) & 0.0651(042) & \\
    D30.48   &      &    & 0.0030 & 0.0978(04) & 0.1296(15) & 0.0860(046) & \\
    \hline
    D45.32sc & 2.10 & 32 & 0.0045 & 0.1198(05) & 0.1480(09) & 0.0886(095) & \\
    \hline\hline
    \vspace*{0.1cm}
  \end{tabular*}
  \caption{Determination of the charged $\mpc$, neutral connected
    $\mpnc$ and neutral $\mpn$ pseudoscalar meson masses from
    simulations with $\nftoo$ flavours of Wilson twisted mass fermions
    at maximal twist and the Iwasaki gauge action. The value of the
    Sommer scale $r_0$ is determined in the chiral limit. The ensemble
    names containing an ``s'' or a ``c'' refer to a change of the strange
    or the charm quark mass, respectively. The values for $\mpc$ agree
    within errors with the once published already in Ref.~\cite{Baron:2010bv}.}
  \label{tab:nf211}
\end{table}
%%%%%%%%%%%%%%%%%%%%%%%%%%%%%%%%%%
The mass-splittings in eqs.~(\ref{eq:w8ms})-(\ref{eq:w6ms}), which are
directly proportional to $\wep$ and $\wsp$ respectively, are
illustrated in Fig.~\ref{fig:MS_con_211}. We observe that the
conditions $\wep <0$ and $\wsp>0$ are fulfilled by the lattice data,
in agreement with the bounds derived in
Refs.~\cite{Damgaard:2010cz,Hansen:2011kk,Kieburg:2012fw}.  In order
to quote the values of the Wilson LECs $W'_{6,8}$, the systematic
effects from quark-mass dependence, residual lattice artifacts and
finite volume effects have to be addressed.
%%%%%%%%%%%%%%%%%%%%%%%%%%%%%%%%%%
%%%%%%%%%%%%%%%%%%%%%%%%%%%%%%%%%%

In the ``large cut-off effects'' power counting and at LO in the
W$\chi$PT chiral Lagrangian, the mass-splittings in
eqs.~(\ref{eq:w8ms})-(\ref{eq:w6ms}) are expected to be independent of
the lattice spacing and the light-quark mass. The possible presence
of such effects might thus signal effects entering at higher orders in
the W$\chi$PT chiral expansion. Since the NLO expressions for these
mass-splittings is currently not available in the literature, we rely in our
systematic error analysis on a separate study of (i) the
continuum-limit of the mass-splittings at a reference mass and (ii)
the comparison of a constant and a linear chiral extrapolation in
$\mpc^2$.

\begin{figure}[t]
  \centering
  \subfigure[\label{fig:MS_a_c2_211a}]{
    \includegraphics[height=6.4cm]{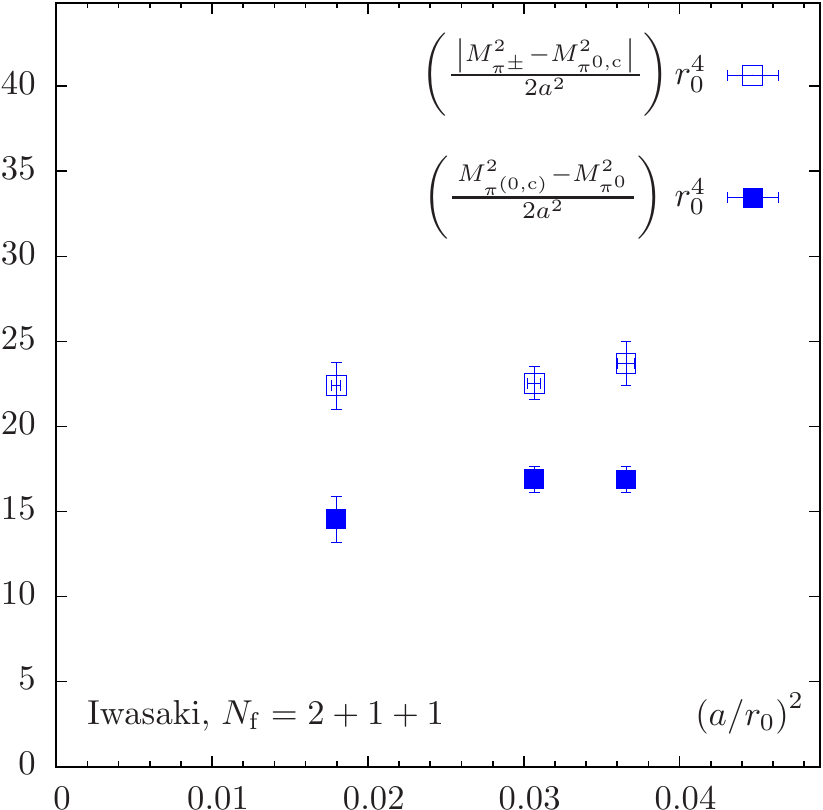}
  }
  \subfigure[\label{fig:MS_a_c2_211b}]{
    \includegraphics[height=6.4cm]{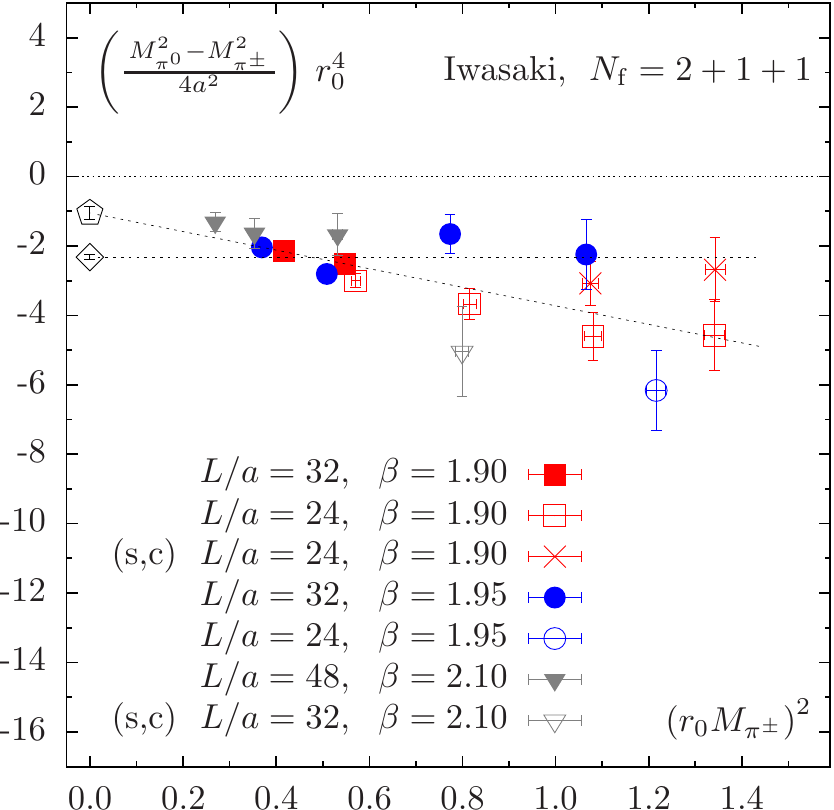}
  }
  \caption{(a) Lattice spacing dependence of the mass-splittings
    $|\mpc^2 - \mpnc^2|/a^2$ -- empty symbols -- and $(\mpnc^2 -
    \mpn^2)/2a^2$ -- filled symbols -- at a reference mass $(\mpc r_0)^2
    \approx 0.55$. For a better visibility, the absolute value of the
    mass splitting is considered in the case of the empty symbols. (b)
    Pion mass-splitting, $(\mpn^2 - \mpc^2)/a^2$, normalised according to
    eq.~(\ref{eq:c2ms}) in order to relate it to the combination of
    Wilson LECs $c_2$. The chiral extrapolation using a constant and a
    linear fit in $\mpc^2$ is also shown. The lattice setup with the
    Iwasaki gauge action and $\nftoo$ flavours of Wilson twisted mass
    fermions is considered.}
  \label{fig:MS_a_c2_211}
\end{figure}

Starting with point (i), we show in Fig.~\ref{fig:MS_con_211} the mass
splittings eq.~(\ref{eq:w8ms}) and (\ref{eq:w6ms}) relevant for $\wep$
and $\wsp$, respectively, as a function of $\mpc^2$.
We first observe that data points with similar values of
$(\mpc r_0)^2$, but coming from different lattice spacings, tend to be
compatible with each other, in particular when considering the larger
lattice sizes, represented by the filled symbols. The lattice spacing
dependence of the two aforementioned mass splittings at a reference
mass $(\mpc r_0)^2 \approx 0.55$ is 
illustrated in Fig.\ref{fig:MS_a_c2_211a}. Note that only the largest
lattice sizes $L$ are considered in this figure. Although the lattice
size slightly varies when changing $\beta$, the lattice data fulfils
$L \gtrsim 2.5$\,fm and $\mpc L \gtrsim 4$ and, therefore, we do not
expect a large effect from a small mismatch in the physical
volume. Fig.\ref{fig:MS_a_c2_211a} suggests that the residual lattice
spacing effects are small. We remind that these lattice artifacts
appear beyond the leading O($a^2$) effects. As an estimate of these
effects, we include in our systematic error analysis the difference
between the values of the mass-splittings from the two finer lattice
spacings.
%%%%%%%%%%%%%%%%%%%%%%%%%%%%%%%%%%
%%%%%%%%%%%%%%%%%%%%%%%%%%%%%%%%%%

Concerning the light-quark mass dependence in point (ii), we can
expect that the mass terms appearing at NLO can contain a linear term
in $\mpc^2$ but also a term of the form $\mpc^2 \log(\mpc^2)$. Indeed,
such terms are present in the W$\chi$PT expression of the mass
splitting between the charged and neutral pion masses at
NLO~\cite{Bar:2010jk}. Since the precise form of the these logarithmic
terms is yet unknown for the mass-splittings considered here, we limit
ourselves to a linear chiral extrapolation in $\mpc^2$. Note that our
data is not precise enough to disentangle possible
logarithmic contributions. We take as our central values the linearly
extrapolated mass-splittings and use the difference with respect to
the constant fit as an estimate of the systematic error from the
light-quark mass dependence.

Finite volume effects are taken into account by adding to the
systematic error the difference between the values of the mass
splittings from two ensembles -- A40.24 and A40.32 -- differing only
by a change of lattice size from $L \approx 2.1$\,fm to
$2.8$\,fm. This is expected to be a conservative choice since these
ensembles correspond to a rather small light-quark mass -- and
therefore finite size effects can be non-negligible. Also, these
ensembles were obtained at the coarsest lattice spacing, where
possible finite size effects (FSE) from the neutral pion mass should
be larger.

As already mentioned, the determination of the Wilson LECs $W'_{6,8}$
from lattice data with $\nftoo$ flavours assumes that the strange and
the charm sea-quarks decouple sufficiently from the light-quark
dynamics. The residual heavy quark mass dependence present in
$W'_{6,8}$ can be studied by varying the strange and charm quark
masses in the neighbourhood of their physical values. This effect is
illustrated in Fig.~\ref{fig:MS_con_211} by the points labelled by
``(s,c)'' in the legend. We use the difference between the
mass-splittings from ensembles A80.24 and A80.24s -- which only differ
by a change of the strange and charm quark masses -- to estimate
this systematic effect. We expect that this is a conservative choice
because (a) the ensemble A80.24s has a strange quark mass which is
very close to the physical point, (b) the change in the strange quark
mass is largest for A80.24 and A80.24s and (c) the effect of strange
sea-quarks should be larger than that of charm quarks.

After combining the previously discussed systematic uncertainties in
quadrature, we obtain the following values for the mass-splittings 
for the case of a lattice setup with $\nftoo$ Wtm fermions and the
Iwasaki gauge action,
%%%%%%%%%%%%%%%%%%%%%%%%%%%%%%%%%%
\begin{eqnarray}
  \left( \frac{ \mpc^2  - \mpnc^2 }{ a^2  } \right) r_0^4\  &=&~-23.0
  \pm 0.7 \pm 3.0\,, \label{eq:vw8msnf211}\\
  \left( \frac{ \mpnc^2 - \mpn^2  }{ 2a^2 } \right) r_0^4\  &=&~+13.8
  \pm 0.6 \pm 5.6\,, \label{eq:vw6msnf211}
\end{eqnarray}
%%%%%%%%%%%%%%%%%%%%%%%%%%%%%%%%%%
where the first error is statistical and the second systematic. The
corresponding values of the Wilson LECs are collected in
Tab.~\ref{tab:WLECnf211}. As already anticipated, the results for
$w'_{6,8}$ are precise enough to identify a definite sign for these
LECs.
%%%%%%%%%%%%%%%%%%%%%%%%%%%%%%%%%%
\begin{table}[t!]
  \centering
  \begin{tabular*}{0.8\textwidth}{@{\extracolsep{\fill}}cccc}
    \hline\hline
          & $w'_8\, r_0^4$ & $w'_8$                    & $\wep\, (r_0^6 W_0^2)$\\
    syst. &  -2.9(4)      & $-[571(32)\,{\rm MeV}]^4$ & -0.0138(22)\\
    \hline
          & $w'_6\, r_0^4$ & $w'_6$                    & $\wsp\, (r_0^6 W_0^2)$\\
    syst. &  +1.7(7)      & $+[502(58)\,{\rm MeV}]^4$ & +0.0082(34)\\
    \hline
          & $c_2\,r_0^4$   & $c_2$                     & $-2\, (2\wsp+\wep)\, (r_0^6 W_0^2)$\\
    lin.  &  -1.1(2)      & $-[444(28)\,{\rm MeV}]^4$ & -0.0050(10)\\ 
    cst.  &  -2.3(1)      & $-[541(24)\,{\rm MeV}]^4$ & -0.0111(10)\\
    \hline\hline
    \vspace*{0.1cm}
  \end{tabular*}
  \caption{Determination of the Wilson LECs $W'_{6,8}$ ($w'_{6,8}$)
    and $c_2$ from a lattice setup with $\nftoo$ Wtm fermions and the
    Iwasaki gauge action. For the values quoted in physical units, the
    input $r_0=0.45(2)$\,fm has been used. The values in the last
    column derive from eq.~(\ref{eq:wk}), where the input $r_0
    f=0.276(12)$ from~\cite{Baron:2010bv} has been used. For the case
    of $W'_{6,8}$ the systematic error analysis described in the text
    has been incorporated in the overall uncertainty indicated by the
    label ``syst.'' in the table. For $c_2$, in the last block, we
    quote separately the results of a constant and a linear chiral
    extrapolation in $\mpc^2$ -- as illustrated in
    Fig.\ref{fig:MS_a_c2_211b} -- and the quoted errors are purely
    statistical.}
  \label{tab:WLECnf211}
\end{table}
%%%%%%%%%%%%%%%%%%%%%%%%%%%%%%%%%%

The combination of LECs $c_2$, can be determined directly from the
mass-splitting $\mpn^2 - \mpc^2$ as indicated in
eq.~(\ref{eq:c2ms}). The measurements of this mass-splitting are
illustrated in Fig.~\ref{fig:MS_a_c2_211b} where the results of a
chiral extrapolation by using a constant and a linear fit in $\mpc^2$
are also shown. For the case of $c_2$, we provide in
Tab.~\ref{tab:WLECnf211} the results from both these chiral
extrapolations and quote in the individual numbers only the
statistical error. The values arising from these extrapolations are
both compatible with a negative sign of $c_2$.

The W$\chi$PT expressions at NLO relevant for $c_2$ have been derived
in Ref~\cite{Bar:2010jk}. In addition to the Wilson LECs appearing at
LO and to the usual Gasser-Leutwyler LECs, other parameters also
appear at NLO. A complete determination of these parameters is beyond
the scope of this study. We postpone such an analysis to a future
dedicated study of the W$\chi$PT description of both the pion mass and
decay constant. First results for the Gasser-Leutwyler LECs, from fits
based on continuum $\chi$PT, have been presented in
Refs.~\cite{Baron:2010bv,Baron:2011sf}.

The connected neutral pion mass, $\mpnc$, is an important ingredient
in order to isolate the individual values of the LECs $W'_{6,8}$. As
already pointed out, the connected neutral pion can be interpreted as
the pion of a mixed action with OS fermions. Such a mixed action has
been used to determine observables in the Kaon
sector~\cite{Constantinou:2010qv,Farchioni:2010tb,Herdoiza:2011gp}. We
note that extensions of the analytical expressions to SU(3) W$\chi$PT
is currently not available in the literature. Contrary to the pion
case, the absence of disconnected diagrams in correlations functions
in the Kaon sector could possibly allow to consider quantities from
which the Wilson LECs can be determined with good accuracy.

%%%%%%%%%%%%%%%%%%%%%%%%%%%%%%%%%%%%%%%%%%%%%%%%%%%%%%
%%%%%%%%%%%%%%%%%%%%%%%%%%%%%%%%%%%%%%%%%%%%%%%%%%%%%%

\subsection{$\nft$ Wtm fermions with tlSym gauge action}

In this section, we again determine the LECs $W'_{6,8}$, but this time
using $\nft$ flavours of Wilson twisted mass fermions and the
tree-level Symanzik improved gauge
action~\cite{Boucaud:2007uk,Boucaud:2008xu,Baron:2009wt}. We already
anticipate that a smaller set of ensembles and of measurements of the
relevant pion masses are available in this case, in comparison to the
$\nftoo$ case discussed previously. Therefore, the resulting
determinations and comparisons might suffer from insufficient control
of systematic effects. However, we think that already a qualitative
comparison can provide useful information to parametrise the size of
cutoff effects from different lattice setups.

%%%%%%%%%%%%%%%%%%%%%%%%%%%%%%%%%%
\begin{table}[t!]
  \centering
  \begin{tabular*}{1.0\textwidth}{@{\extracolsep{\fill}}ccccccc}
    \hline\hline
    $\beta$ & $L/a$ & $a\mu_\ell$ & $a\mpc$ & $a\mpnc$ & $a\mpn$ & $r_0/a$\\
    \hline
    3.90 & 32 & 0.0040 & 0.1338(02) & 0.2080(30) & 0.1100(080) & 5.35(4)\\
    & 24 & 0.0040 & 0.1362(07) & 0.2120(30) & 0.1090(070) & \\
    & 16 & 0.0040 & 0.1596(30) & 0.2226(95) & -           & \\
    & 24 & 0.0064 & 0.1694(04) & -          & 0.1340(100) & \\
    &    & 0.0085 & 0.1940(05) & -          & 0.1690(110) & \\
    & 16 & 0.0074 & 0.1963(17) & 0.2541(55) & -           & \\
    \hline
    4.05 & 32 & 0.0030 & 0.1038(06) & 0.1500(30) & 0.0900(060) & 6.71(4)\\
    & 20 & 0.0030 & 0.1191(41) & 0.1571(62) & -           & \\
    & 32 & 0.0060 & 0.1432(06) & 0.1800(20) & 0.1230(060) & \\
    \hline
    4.20 & 24 & 0.0020 & 0.0941(31) & 0.1157(61) & -           & 8.36(6)\\
    \hline\hline
    \vspace*{0.1cm}
  \end{tabular*}
  \caption{Determination of the charged $\mpc$, neutral connected
    $\mpnc$ and neutral $\mpn$ pseudoscalar meson masses from
    simulations with $\nft$ flavours of Wilson twisted mass fermions
    at maximal twist and the tree-level Symanzik improved gauge
    action~\cite{Baron:2009wt}. The value of the Sommer scale $r_0$ is
    determined in the chiral limit~\cite{Blossier:2010cr}.}
  \label{tab:nf2}
\end{table}
%%%%%%%%%%%%%%%%%%%%%%%%%%%%%%%%%%

The simulations considered in this
work~\cite{Baron:2009wt,Cichy:2010ta} were performed at three values
of the lattice gauge coupling $\beta =3.90$, $4.05$ and $\beta=4.20$,
corresponding to values of the lattice spacing $a\approx 0.08$\,fm,
$0.07$\,fm and $0.05$\,fm, respectively. The charged pion mass $\mpc$
approximately ranges from $310$\,MeV to $460$\,MeV. Physical spatial
volumes range from $(1.3\,\mathrm{fm})^3$ to $(2.6\,\mathrm{fm})^3$
and ensembles which differ only by the lattice size have been
considered in order to address the size of finite volume effects in
the determination of the Wilson LECs.

\begin{figure}[t!]
  \centering \subfigure[\label{fig:MS_con_2a}]{
    \includegraphics[height=6.5cm]{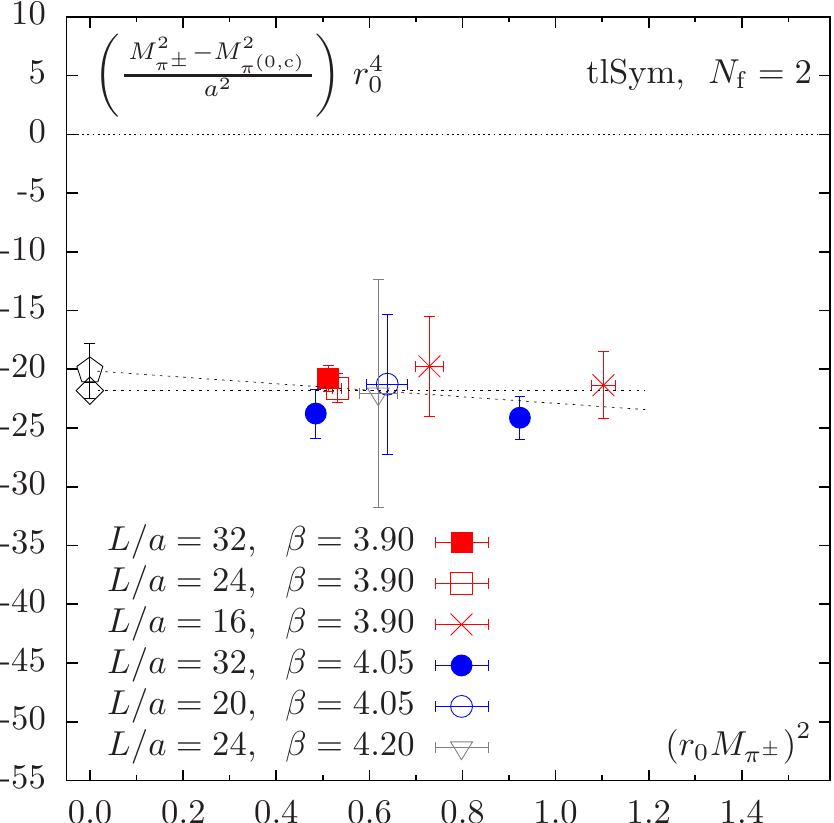}
  } \subfigure[\label{fig:MS_con_2b}]{
    \includegraphics[height=6.5cm]{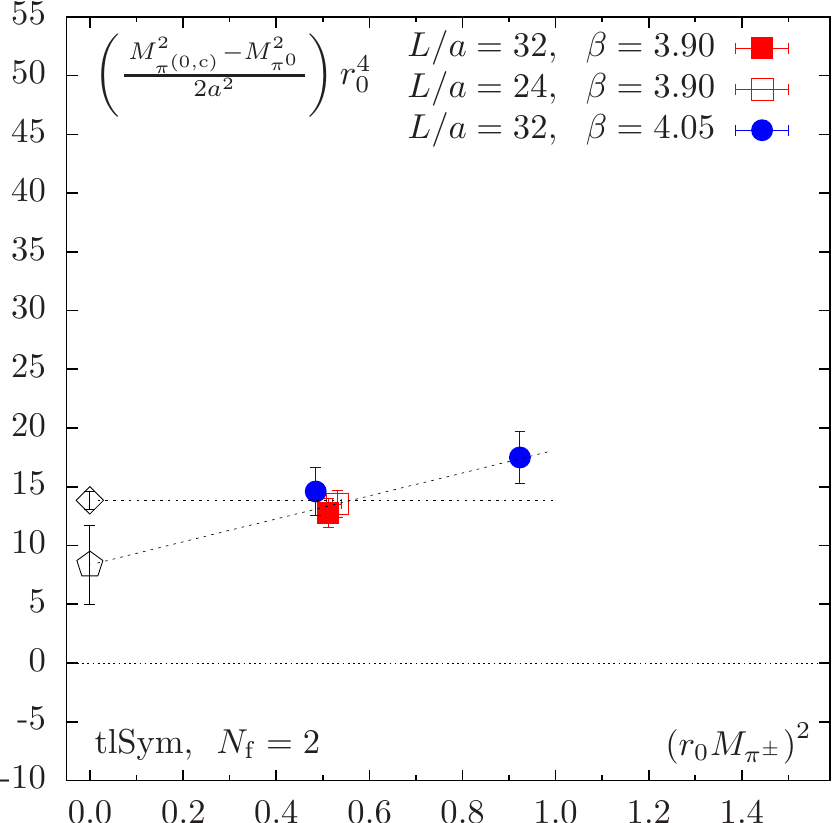}
  }
  \caption{Determination of the mass-splittings (a) $(\mpc^2 -
    \mpnc^2)/a^2$ in eq.~(\ref{eq:w8ms}) and (b) $(\mpnc^2 -
    \mpn^2)/2a^2$ in eq.~(\ref{eq:w6ms}) as a function of
    $\mpc^2$. These quantities are in units of the chirally
    extrapolated Sommer scale $r_0$. The mass-splittings are directly
    related to the Wilson LECs $\wep$ and $\wsp$. The lattice setup
    with the tree-level Symanzik improved (tlSym) gauge action and
    $\nft$ flavours of Wilson twisted mass fermions is
    considered. Filled and empty symbols signal a change in the
    lattice size.}
  \label{fig:MS_con_2}
\end{figure}

\begin{figure}[t]
  \centering
  \subfigure[\label{fig:MS_L_c2_2a}]{
    \includegraphics[height=6.4cm]{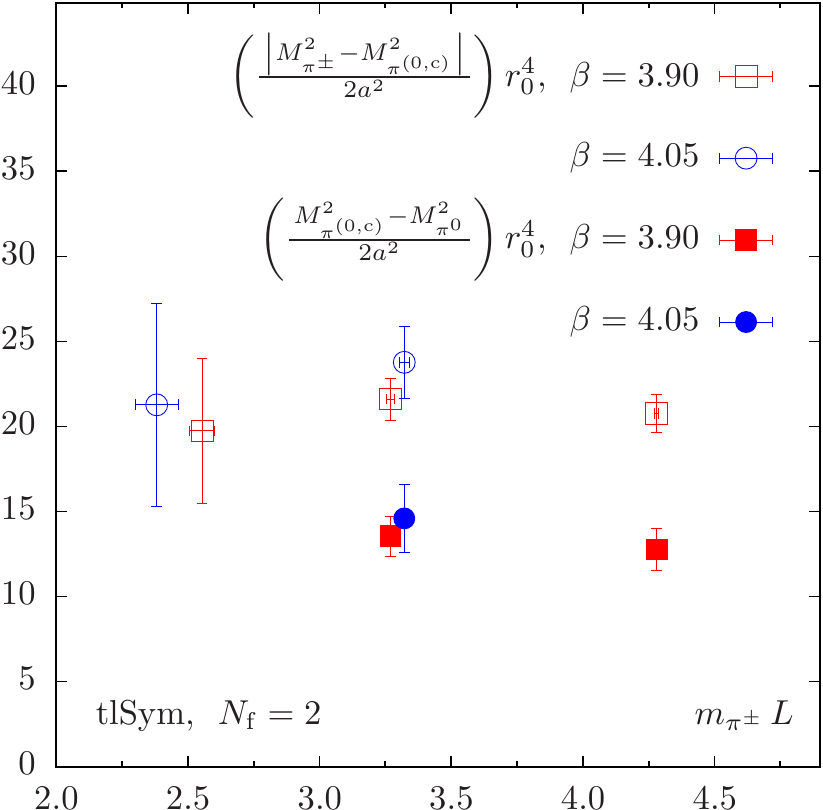}
  }
  \subfigure[\label{fig:MS_L_c2_2b}]{
    \includegraphics[height=6.4cm]{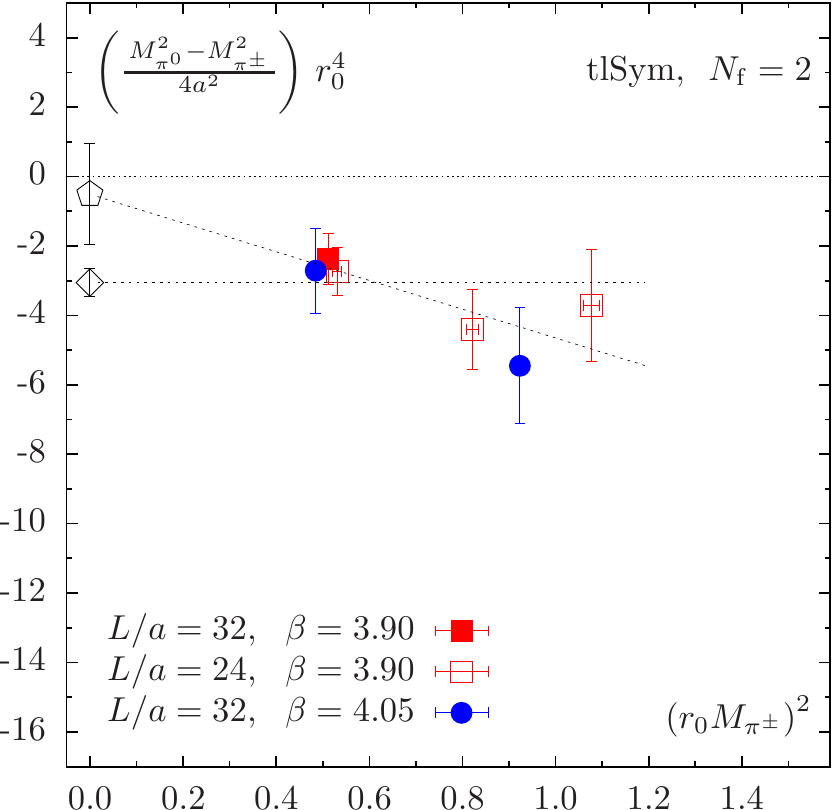}
  }
  \caption{(a) Finite volume effects on the mass-splittings $|\mpc^2 -
    \mpnc^2|/a^2$ -- empty symbols -- and $(\mpnc^2 - \mpn^2)/2a^2$ --
    filled symbols. Note that the absolute value of the mass-splitting
    is used in the case of empty symbols. (b) Pion mass-splitting,
    $(\mpn^2 - \mpc^2)/a^2$, normalised according to
    eq.~(\ref{eq:c2ms}) in order to relate it to the combination of
    Wilson LECs $c_2$. The chiral extrapolation using a constant and a
    linear fit in $\mpc^2$ is also shown. The lattice setup with the
    tree-level Symanzik improved (tlSym) gauge action and $\nft$
    flavours of Wilson twisted mass fermions is considered. }
  \label{fig:MS_L_c2_2}
\end{figure}

The values of the pseudoscalar meson masses~\cite{Baron:2009wt} for
the $\nft$ ensembles are collected in Table~\ref{tab:nf2}. The
mass-splittings in eqs.~(\ref{eq:w8ms})-(\ref{eq:w6ms}) are
illustrated in Fig.~\ref{fig:MS_con_2}.
%%%%%%%%%%%%%%%%%%%%%%%%%%%%%%%%%%
%%%%%%%%%%%%%%%%%%%%%%%%%%%%%%%%%%
In order to explore the systematic effects present in these
determinations, we follow a similar path to that described for the
case of $\nftoo$ ensembles. The availability of ensembles differing
only by the physical volume allows to address the size of FSE in the
mass-splittings. This is illustrated in Fig.~\ref{fig:MS_L_c2_2a}. At
$\beta=3.90$, a set of three ensembles with $L/a=16,~24$ and $32$
could be used for the case of $(\mpc^2 - \mpnc^2)/a^2$. With the
current statistical uncertainties, no clear signs of FSE can be
observed in the data. Furthermore, data from two different lattice
spacings -- $\beta=3.90$ and $4.05$ -- agree within errors, indicating
the absence of large residual lattice artifacts in these
mass-splittings.
%%%%%%%%%%%%%%%%%%%%%%%%%%%%%%%%%%
%%%%%%%%%%%%%%%%%%%%%%%%%%%%%%%%%%
However, the lack of sufficient data does not allow to address the
mass dependence of the mass-splitting $(\mpnc^2 - \mpn^2)/a^2$ in a
satisfactory way. In analogy to the $\nftoo$ case, we include the
deviation between a constant and a linear extrapolation in $\mpc^2$ in
the estimate of the systematic uncertainties. The central value is
taken from the result of the linear fit.

For the case of a lattice setup with $\nft$ Wtm fermions at maximal
twist and the tlSym gauge action, we obtain the following values for
the mass-splittings
%%%%%%%%%%%%%%%%%%%%%%%%%%%%%%%%%%
\begin{eqnarray}
  \left( \frac{ \mpc^2  - \mpnc^2 }{ a^2  } \right) r_0^4\  &=&~-20.1
  \pm 2.3 \pm 1.7\,, \label{eq:vw8msnf2}\\
  \left( \frac{ \mpnc^2 - \mpn^2  }{ 2a^2 } \right) r_0^4\  &=&~~+8.4
  \pm 3.3 \pm 5.5\,, \label{eq:vw6msnf2}
\end{eqnarray}
%%%%%%%%%%%%%%%%%%%%%%%%%%%%%%%%%%
where the first error is statistical and the second systematic. The
corresponding values of the Wilson LECs are collected in
Tab.~\ref{tab:WLECnf2}.
%%%%%%%%%%%%%%%%%%%%%%%%%%%%%%%%%%
\begin{table}[t!]
  \centering
  \begin{tabular*}{0.6\textwidth}{@{\extracolsep{\fill}}cccc}
    \hline\hline
    $w'_8\, r_0^4$ & $w'_8$                     & $\wep\, (r_0^6 W_0^2)$\\
    -2.5(4)       & $-[552(025)\,{\rm MeV}]^4$ & -0.0119(17)\\
    \hline
    $w'_6\, r_0^4$ & $w'_6$                     & $\wsp\, (r_0^6 W_0^2)$\\
    +1.0(8)       & $+[443(138)\,{\rm MeV}]^4$ & +0.0049(38)\\
    \hline\hline
    \vspace*{0.1cm}
  \end{tabular*}
  \caption{Determination of the Wilson LECs $W'_{6,8}$ ($w'_{6,8}$)
    from a lattice setup with $\nft$ Wtm fermions and the tlSym gauge
    action. For the values quoted in physical units, the input
    $r_0=0.45(2)$\,fm has been used. The values in the last column
    derive from eq.~(\ref{eq:wk}), where the input $r_0 f=0.275(6)$
    from~\cite{Baron:2009wt} has been used.}
  \label{tab:WLECnf2}
\end{table}
%%%%%%%%%%%%%%%%%%%%%%%%%%%%%%%%%%
The LEC $W'_{8}$ has recently been determined from a mixed action
involving the same $\nft$ lattice action in the sea sector as that
described here, but with Neuberger overlap valence
quark~\cite{Cichy:2012vg}. The value quoted in
Ref.~\cite{Cichy:2012vg}, $\wep\, (r_0^6 W_0^2)=-0.0064(24)$, differs
from the estimate in Tab.~\ref{tab:WLECnf2} at the 2-sigma
level. However, we stress once more that our present $\nft$
estimate does not include a complete assessment of the systematic
errors.

The determination of $c_2$ from $\nft$ ensembles is illustrated in
Fig.\ref{fig:MS_L_c2_2b}. More data would be needed to isolate the
residual mass-dependence present in $c_2$. For this reason we opt for
quoting separately the results of a constant and a linear chiral
extrapolation in $\mpc^2$,
%%%%%%%%%%%%%%%%%%%%%%%%%%%%%%%%%%
\begin{eqnarray}
  c_2\,r_0^4\,[\,{\rm cst.}\,]\  &=&~-3.1
  \pm 0.4\,, \label{eq:c2cstnf2}\\
  c_2\,r_0^4\,[\,{\rm lin.}\,]\  &=&~-0.5
  \pm 1.5\,, \label{eq:c2linnf2}
\end{eqnarray}
%%%%%%%%%%%%%%%%%%%%%%%%%%%%%%%%%%
where the errors are statistical only. These values are consistent
with those arising from the measurements of pseudoscalar meson masses
in Ref.~\cite{Baron:2009wt} and are also very similar to the result
obtained in Ref.~\cite{Colangelo:2010cu} for $K=-4c_2$ from twisted
mass finite volume effects. All the lattice measurements favour a
negative sign of $c_2$. However, as already mentioned, more data would
be needed to properly address the residual mass dependence.

We refer to Refs.~\cite{Baron:2009wt,Blossier:2010cr} for more details
about the description of the pion mass and decay constant by means of
$\chi$PT expressions including discretisation effects.

%%%%%%%%%%%%%%%%%%%%%%%%%%%%%%%%%%%%%%%%%%%%%%%%%%%%%%%%%%%%%%%%%%%%%%%%%%%%%%
%%%%%%%%%%%%%%%%%%%%%%%%%%%%%%%%%%%%%%%%%%%%%%%%%%%%%%%%%%%%%%%%%%%%%%%%%%%%%%

\section{Discussion}
\label{sec:discussion}

In this section we collect a few comments concerning the extraction of
the Wilson LECs.  The LECs $\wsp$ and $c_2$ depend on the neutral pion
mass $\mpn$. The statistical error in $\mpn$ is dominated by the
contribution of disconnected diagrams.\footnote{It is interesting to
  note that Table~\ref{tab:nf211} indicates that the relative error on
  the neutral pion mass is roughly independent of the light-quark mass
  and that it decreases when increasing the volume. In practice, in
  the current simulation conditions, this implies that the
  measurements of $\mpn$ are statistically more precise for the
  ensembles with lighter quark masses.} We observe that the precise
form of the light-quark mass dependence of the mass-splittings related
to $\wsp$ and $c_2$ -- see e.g. Figs.~\ref{fig:MS_con_211b} and
\ref{fig:MS_a_c2_211b} -- cannot be addressed within the present
uncertainties. As previously discussed this mass dependence can arise
at NLO in the W$\chi$PT expansion. We can, however, not exclude at this
stage that this higher-order effects are negligible. This issue is
particularly relevant for the case of $c_2 \propto -(2 w'_6+ w'_8)$,
where a partial cancellation of the effect of $w'_6$ and $w'_8$ is
present.  The presence of higher order effects in the determination of
$c_2$ has also been discussed in Ref.~\cite{Bernardoni:2011fx}.  We
recall that the sign of $c_2$ controls the appearance of an Aoki ($c_2
> 0$) or of a Sharpe-Singleton ($c_2 < 0$) scenario for the phase
structure of Wilson fermions.

From the previously discussed determinations, a comparison of the
values of the Wilson LECs from the lattice actions (i) $\nftoo$
flavours of Wtm fermions at maximal twist and Iwasaki gauge action and
(ii) $\nft$ flavours of Wtm quarks and tlSym gauge action, suggests
that $W'_{6,8}$ and $c_2$ do not vary significantly in between
these two setups.  Let us extend this observation by performing a
comparison of the pion mass-splittings as determined from different
lattice actions. We stress that a complete assessment of the overall
uncertainty is not available for most of these measurements and
therefore the comparison remains at the qualitative level.

\begin{figure}[t]
  \centering
  \subfigure[\label{fig:comparisona}]{
    \includegraphics[height=7.8cm]{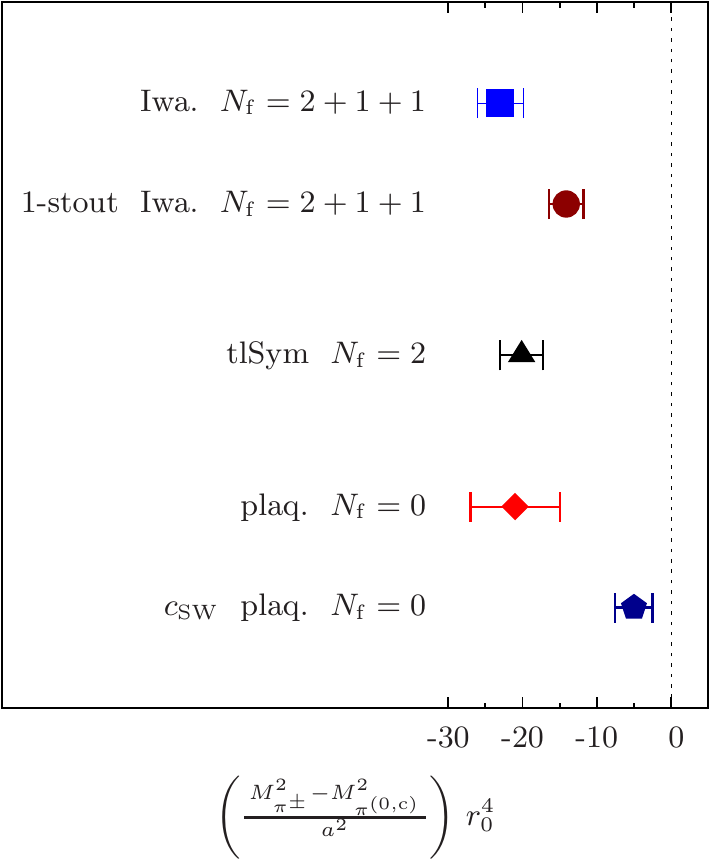}
  }
  \subfigure[\label{fig:comparisonb}]{
    \includegraphics[height=7.8cm]{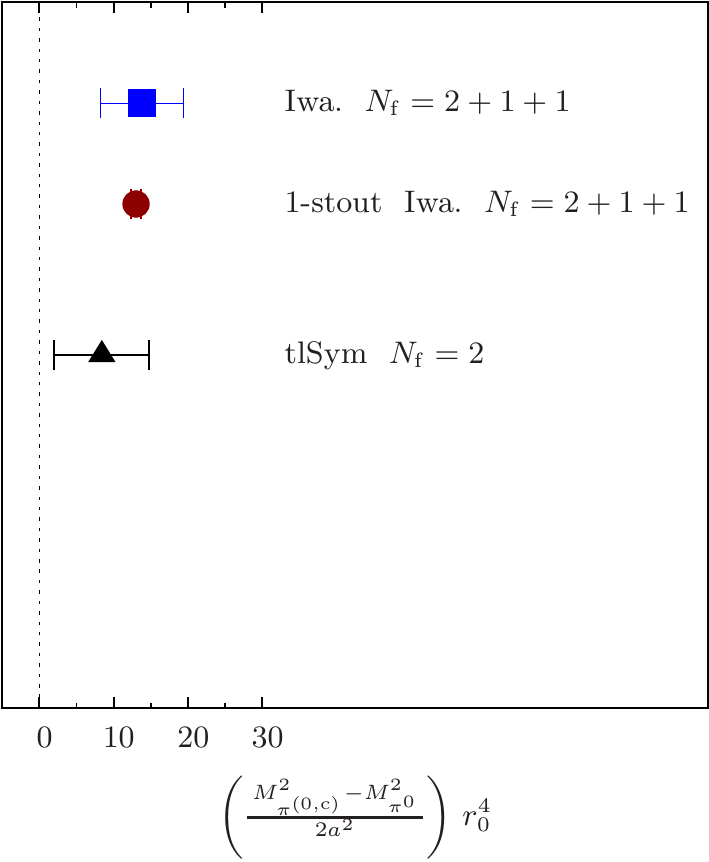}
  }
  \caption{Comparison of the values of the mass-splittings (a)
    $(\mpc^2 - \mpnc^2)/a^2$ and (b) $(\mpnc^2 - \mpn^2)/2a^2$ from
    different lattice setups (see text for details).}
  \label{fig:comparison}
\end{figure}

Lattice simulations with $\nftoo$ Wtm fermions and the Iwasaki gauge
action, but including in addition one iteration of stout smearing --
labelled 1-stout -- have been reported in Ref.~\cite{Baron:2010bv}. A
qualitative comparison of the effect of the stout smearing on the size
of the mass-splittings is shown in Fig.~\ref{fig:comparison}. Note
that a single ensemble is used in the estimate of the mass-splittings
for the case of stout smearing. The used smearing seems to help in
reducing the magnitude of the splitting $(\mpc^2 - \mpnc^2)/a^2$ while
-- given the current uncertainties --it does not introduce a
significant change in $(\mpnc^2 - \mpn^2)/a^2$.
%%%%%%%%%%%%%%%%%%%%%%%%%%%%%%%%%%
%%%%%%%%%%%%%%%%%%%%%%%%%%%%%%%%%%%%%%%%%%%%%%%%%%%%%%%%%%%%%%%%%%%%%%%%%%%%%%

Fig.~\ref{fig:comparison}(a) also includes the estimates of the
mass-splitting $(\mpc^2 - \mpnc^2)/a^2$ as determined from quenched
ensembles either with or without the presence of the
Sheikholeslami-Wohlert term. Maximally twisted-mass fermions and the
plaquette gauge action are used in both cases. The values of the
mass-splitting are derived from studies available in the
literature. For the case without the Sheikholeslami-Wohlert term,
results are taken from Ref.~\cite{Jansen:2005cg}. We use the
information from different lattice spacings and quark masses to
estimate the size of the systematic effects in $(\mpc^2 -
\mpnc^2)/a^2$. For the case in which the Sheikholeslami-Wohlert term
is included -- labelled $c_{\rm SW}$ -- we follow
Ref.~\cite{Dimopoulos:2009es}, where the non-perturbative
determination of $c_{\rm SW}$ was used.

For cases other than those involving the Sheikholeslami-Wohlert term
or stout-smearing, Fig.~\ref{fig:comparisona} suggests that, with the
current precision, the value of the mass-splitting $(\mpc^2 -
\mpnc^2)/a^2$ does not significantly depend on a simultaneous change
of the number of flavours $N_{\rm f}$ and of the gauge action. Note
however that we cannot exclude that a change only in $N_{\rm f}$ or
only in the parameter $b_1$ of the gauge action, leads to a different
conclusion. It would be very desirable, if further actions are
investigated and the precision could be increased.

One important observation arising from the measurements in
Ref.~\cite{Dimopoulos:2009es} is that the introduction of the
Sheikholeslami-Wohlert term in quenched studies significantly reduces
the size of O($a^2$) effects by lowering the value of the
mass-splitting $(\mpc^2 - \mpnc^2)/a^2$. It would thus be interesting
to study whether this result still holds for simulations with
dynamical fermions and whether the value of $(\mpnc^2 - \mpn^2)/2a^2$
is also reduced in that case.

%%%%%%%%%%%%%%%%%%%%%%%%%%%%%%%%%%%%%%%%%%%%%%%%%%%%%%%%%%%%%%%%%%%%%%%%%%%%%%
%%%%%%%%%%%%%%%%%%%%%%%%%%%%%%%%%%%%%%%%%%%%%%%%%%%%%%%%%%%%%%%%%%%%%%%%%%%%%%

\section*{Conclusions}
\label{sec:concl}

We have presented the determination of the Wilson LECs $W'_{6,8}$ and
$c_2$, pa\-ra\-me\-trising the size of O($a^2$) lattice artifacts in
W$\chi$PT, from simulations with a lattice action composed out of the
Iwasaki gauge action and $\nftoo$ flavours of Wilson twisted mass
fermions at maximal twist. The values of $W'_{6,8}$ include a rather
complete account of the systematic uncertainties. Our measurements
satisfy the recently derived
bounds~\cite{Damgaard:2010cz,Hansen:2011kk,Kieburg:2012fw}, $\wep <0$
and $\wsp > 0$.

We have also explored the dependence of the mass-splittings $(\mpc^2 -
\mpnc^2)/a^2$ and $(\mpnc^2 - \mpn^2)/2a^2$ on the choice of the
lattice action. From this qualitative comparison, it is tempting to
conjecture that a lattice action made of dynamical twisted mass
fermions including the Sheikholeslami-Wohlert and smearing might lead
to a reduction of the mass-splitting $(\mpc^2 -
\mpnc^2)/a^2$. Further studies are needed to clarify this point and to
extend it to the case of $(\mpnc^2 - \mpn^2)/2a^2$.

A partial cancellation of the contributions from $W'_{6}$ and $W'_{8}$
implies that the residual mass-dependence of $c_2$ is more sensitive
to higher order effects in the W$\chi$PT expansion. While this
potential reduction of $c_2$ would certainly be beneficial, the
precise determination of its value cannot be achieved with the
currently available data.

The determination of the Wilson LECs can help to quantify the size of
O($a^2$) terms in a given lattice action. Knowing these LECs, in
particular with better precision, can significantly contribute to
design a lattice fermion action with small lattice artifacts, thus
allowing to reach the continuum limit in a better controlled way. In
addition, an independent calculation of the Wilson LECs, as carried
through here, can be used in chiral perturbation theory fits of light
meson observables by constraining these fits to lattice data. Thus we
think that the study of the (connected and full) neutral and charged
pion masses of this work can be beneficial for many other groups
working with Wilson-like lattice fermions.

%%%%%%%%%%%%%%%%%%%%%%%%%%%%%%%%%%%%%%%%%%%%%%%%%%%%%%%%%%%%%%%%%%%%%%%%%%%%%%
%%%%%%%%%%%%%%%%%%%%%%%%%%%%%%%%%%%%%%%%%%%%%%%%%%%%%%%%%%%%%%%%%%%%%%%%%%%%%%

\section*{Acknowledgements}

We thank Giancarlo Rossi for useful comments on the manuscript. The
computer time for this project was made available to us by the John
von Neumann-Institute for Computing (NIC) on the JUDGE and Jugene
systems in J{\"u}lich and the IDRIS (CNRS) computing center in
Orsay. In particular we thank U.-G.~Mei{\ss}ner for granting us access
on JUDGE. Falk Zimmermann cross-checked correlation functions for one
of our ensembles, which we gratefully acknowledge.  G. H. acknowledges
the support by DFG (SFB 1044), the Spanish Ministry for Education and
Science project FPA2009-09017, the Consolider-Ingenio 2010 Programme
CPAN (CSD2007-00042), the Comunidad Aut\'onoma de Madrid (HEPHACOS
P-ESP-00346 and HEPHACOS S2009/ESP-1473) and the European project
STRONGnet (PITN-GA-2009-238353). K. J. was supported in part by the
Cyprus Research Promotion Foundation under contract
$\Pi$PO$\Sigma$E$\Lambda$KY$\Sigma$H/EM$\Pi$EIPO$\Sigma$/0311/16.
This work has been supported in part by the DFG
Sonderforschungsbereich/ Transregio SFB/TR9. Two of the authors
(K. O. and C. U.) were supported by the Bonn-Cologne Graduate School
(BCGS) of Physics and Astronomie. This project was supported in parts
by the DFG in SFB/TR16 and CRC 110.

%%%%%%%%%%%%%%%%%%%%%%%%%%%%%%%%%%%%%%%%%%%%%%%%%%%%%%%%%%%%%%%%%%%%%%%%%%%%%%
%%%%%%%%%%%%%%%%%%%%%%%%%%%%%%%%%%%%%%%%%%%%%%%%%%%%%%%%%%%%%%%%%%%%%%%%%%%%%%
\bibliographystyle{h-physrev5}
\bibliography{bibliography}

\end{document}